\documentclass{aa}  

\usepackage{graphicx}
\usepackage{xcolor}
\usepackage{hyperref}
\usepackage{academicons}
\usepackage{txfonts}
\begin{document}

   \title{Validation of the RR Lyrae period determination in the Pan-STARRS PS1 3$\pi$ survey with K2}

%   \subtitle{I. Overviewing the $\kappa$-mechanism}

   \author{Adrienn Forr\'o
          \inst{1,2}
          \and
          L\'aszl\'o Moln\'ar
          \inst{1,2}
          \and
          Emese Plachy
          \inst{1,2}
          \and
          \'Aron Juh\'asz
          \inst{1}
          \and
          R\'obert Szab\'o
          \inst{1,2}
%          C. Ptolemy\inst{2}\fnmsep\thanks{Just to show the usage
%          of the elements in the author field}
          }

   \institute{Konkoly Observatory, HUN-REN Research Centre for Astronomy and Earth Sciences, MTA Centre of Excellence, H-1121, Konkoly Thege Mikl\'os \'ut 15-17, Budapest, Hungary\\
              \email{forro.adrienn@csfk.org}
        %\and
            %CSFK, MTA Centre of Excellence, H-1121 Konkoly Thege Miklós út 15-17, Budapest, Hungary\
         %\and
             %Eötvös Loránd University, H-1117, Budapest, Pázmány Péter sétány 1/a, Hungary
             \and
             Eötvös Loránd University, Institute of Physics and Astronomy, H-1117, Budapest, Pázmány Péter sétány 1/a, Hungary\\            
             }

   \date{Received September 15, 1996; accepted March 16, 1997}

% \abstract{}{}{}{}{} 
% 5 {} token are mandatory
 
  \abstract
   {The Pan-STARRS 3$\pi$ survey has detected hundreds of thousands of variable stars thanks to its coverage and 4-year time span, even though the sampling of the light curves is relatively sparse. These light curves contain only 10-15 detections in each of the five filters (g,r,i,z,y). During the K2 mission, the \textit{Kepler} space telescope observed along the ecliptic plane with a high sampling frequency, although only for about 80 days in each of its campaigns.}
   {Crossmatching and investigating the RR Lyrae stars observed by both K2 and Pan-STARRS can serve as a valuable tool to validate the classification and period determination of the ground-based survey.}
   {We used the Sesar catalogue of RR Lyrae stars detected by Pan-STARRS. After determining the overlap between the stars observed by both Pan-STARRS and K2, we also considered the \textit{Gaia} DR3 SOS RR Lyrae catalogue %\citet{Clementini}
   data for the list of these stars wherever it was available. The frequencies of the light variations were calculated by applying the Lomb-Scargle periodogram method on the K2 light curves that were prepared with autoEAP photometry. The calculated frequencies of the stars then were compared with those given in the Sesar catalogue and the \textit{Gaia} DR3 RR Lyrae catalogue.}
   {We found that for the majority of the stars, the classification (95.6 \%) and the frequency determination (90.1 \%) of the Pan-STARRS RR Lyrae stars were consistent within 0.03 $d^{-1}$ with those that we derived from the K2 autoEAP light curves. For a significant subset of the sample, 7.4 \%, however, an offset of 1 or 2 $d^{-1}$ was found in the frequencies. These are the result of the sampling of the detections, because Pan-STARRS  observations are affected by diurnal cycles, whereas \textit{Kepler} carried out measurements quasi-continuously. We found that RRc subtypes are significantly more affected (25.3\%) than RRab subtypes (3.7\%), which is most likely caused by RRc stars having less sharp light curve features. Validation via space-based data will be important for future ground-based surveys, as well.}
   {}


   \maketitle
%
%-------------------------------------------------------------------

\section{Introduction} \label{intro}

RR Lyrae stars are old, population II, core-helium-burning variable
stars, found on the horizontal branch of the Hertzsprung-Russell diagram. Thanks to their high luminosity and the well-studied period-luminosity relation, they can be used as tracers of the galactic structure of the Milky Way \citep{Catelan2015}. They are commonly found in the Galactic halo, and especially in globular clusters, but they appear in large numbers in the Galactic Bulge, the thick disk, and in a metal-poor thin disk \citep{Iorio2021, D'Orazi2024} as well. They also trace extragalactic structures and can be used to identify close-by dwarf galaxies as well as streams and tidal tails of perturbed galaxies, such as the Sagittarius (Sgr) dwarf galaxy and stream \citep{Ibata}. \

Mapping and statistical studies really took off with the start of extensive sky surveys in the early 2000s such as OGLE (Optical Gravitational Lensing Experiment \citep{OGLE}), SDSS (Sloan Digital Sky Survey \citep{SDSS}), NSVS (Northern Sky Variability Survey \citep{NSVS}), ASAS (All-Sky Automated Survey \citep{ASAS}), or the VVV (VISTA Variables in the Via Lactea \citep{Minniti2017}). Since the identification of the Sagittarius stream, numerous other dwarf galaxies and streams have been discovered as a result of the availability of large amounts of data \citep{Weisz}. In addition,\citet{Drake2013} used light curves from the Catalina Surveys, for example, to analyse RRab stars in the outer halo. \

Pan-STARRS observed 75\% of the sky during its 3$\pi$ survey, which lasted for four years. During this time span, the telescope made photometric measurements in each of the five  $grizy$ SDSS filters \citep{PS}. The number of detections for a given object can vary; on average, there are approximately 60 detections in total, 10-15 in each filter. The sampling is non-uniform, which makes it possible to use the obtained light curves to identify variable objects of different types and frequencies despite the relatively low number of detections. Large sky surveys have proved to be excellent sources of data for identifying and tracking the distribution of variables, such as RR Lyrae stars. For example, \citet{Hernitschek2016} analysed the Pan-STARRS PS1 3$\pi$ data and found $\sim 1$ million quasi-stellar objects (QSOs) and $\sim 150,000$ RR Lyrae candidates, and \citet{Sesar2013} used the re-calibrated LINEAR data of RR Lyrae stars to trace the structure of the galactic halo. It is worth noting, however, that ground-based surveys are limited by the diurnal cycle that affects both sampling and period determination. \

Space-based observations, on the other hand, are less heavily affected by the rotation of the Earth, and therefore this kind of measurements can be an important complement to ground-based photometry data. The \textit{Kepler} space telescope made quasi-continuous observations of a dedicated area of the sky, concentrating on pre-selected target stars in the direction of Lyrae and Cygnus, in order to discover exoplanet transits \citep{Borucki2010, Borucki2016}. These datasets have proved to be invaluable in the field of stellar variability. After the spacecraft was no longer able to continue its original mission as a result of the failing reaction wheels, its second mission, named K2, started, during which the orientation of the telescope was changed to observe in the ecliptic plane \citep{Howell2014}, and a new observing mode had to be developed. Similar to its original mission, the sampling remained frequent; however, these light curves typically span about 80 days (with $\sim$ 600--4000 epochs) in contrast to the previous four years. \

A large number of new variable stars have already been discovered thanks to the robust nature of the survey programs, even though all-sky missions, such as the \textit{Gaia} mission \citep{DR3, Clementini}, could only collect sparse photometric data from each observed star. The light curves of the \textit{Gaia} DR3 RR Lyrae catalogue contain on average 38-39 measurements, with 12 and 257 being the minimum and maximum number of observation points. It can make the identification, the classification, or the period determination challenging. The completeness of the \textit{Gaia} DR3 catalogue was estimated by \citet{Clementini}, who found it to be 79--94 \%, based on the OGLE data, which they considered almost complete. However, they found that comparison with a compilation of existing catalogues outside of the OGLE fields is more complicated. \

%for example \citet{Molnar2018} showed that the completeness of the \textit{Gaia} DR2 RR Lyrae catalog is around 75\%. \

The overlap between the all-sky but sparsely sampled ground-based data from Pan-STARRS, and the frequently sampled space-based measurements from K2 over a limited sky area, provide an excellent opportunity to compare and validate the accuracy of these observations. Therefore, the primary aim of this research is to investigate variable stars, specifically RR~Lyrae stars, through the lens of both Pan-STARRS and K2 in order to be able to compare the accuracy of the identification and the period determination. \

In \textit{Section \ref{methods}}, we describe our methods in detail as well as the data and catalogues that served as a basis for the analysis. The results and conclusions are presented in \textit{Sections \ref{results}} and \textit{\ref{conclusions}}. \

\section{Data and methods} \label{methods}

\subsection{Data and catalogues}

\

There are multiple catalogues that contain RR Lyrae stars observed by Pan-STARRS. For example, in their study, \citet{Hernitschek2016} searched for RR~Lyrae-type variables and QSOs in the light curves, analysing the first 3.5 years of observations of the Pan-STARRS dataset. They applied a machine-learning algorithm (a random forest classifier) to determine the probability of the light curve showing RR Lyrae or QSO-type variability. In the published catalogue, these probability scores can be found, but not the periodicity of a possible variability. Another catalogue is the \citet{Sesar2017} catalogue, which is exclusively dedicated to RR Lyrae stars. They applied a template fitting method and then used a machine-learning-based classification. The catalogue contains not only the classifications of the stars and the respective probability scores but also the determined periods. There is a significant overlap between these two catalogues. Because the goal is the analysis and comparison of not only the classifications but also the periods of the stars, we decided to proceed using the \citet{Sesar2017} catalogue. \

%A classification of RR Lyrae stars, similar to the Pan-STARRS search, was made based on the Dark Energy Survey observations as well \citep{DES}. However, those fields have little to no overlap with the K2 observation, so we omitted them from this work. \

 After the original mission of the \textit{Kepler} space telescope ended, the precision of keeping the position of the targets got worse for K2. Therefore in some cases, it can be challenging to determine the optimal aperture for the photometry in order to obtain the light curve with the best signal-to-noise ratio. We used the autoEAP light curves provided by \citet{EAP2022}. In their work, they automated the process of determining the optimal extended aperture for the photometry.\footnote{\url{https://github.com/konkolyseismolab/autoeap}} \

 \subsection{Methods and analysis} \label{2_2}

Hereafter, when discussing the steps of data reduction, a star observed by Pan-STARRS means that it was identified and listed in the \citet{Sesar2017} catalogue. A star observed by K2 means that it was actually observed in at least one of the campaigns, as is explained below. Some stars are observed in multiple campaigns. We started the study by crossmatching the targets to obtain those stars that were potentially observed by both Pan-STARRS and K2. \

To determine the overlap, we used the K2fov python tool \citep{k2fov} to find those stars from the Pan-STARRS RR Lyrae catalogues that fall onto the field of view of the K2 mission in at least one of the campaigns. However, the mere fact that a star's position falls onto the K2 field of views does not guarantee that it was actually included in the observations. Only the pixels around the pre-selected targets were downloaded due to bandwidth constraints. For this reason, after we obtained the primary list of crossmatched stars, we also checked the EPIC catalogue \citep{Huber2016} to see whether the selected stars were observed. We focused on those stars that were among the main targets of the K2 mission in order to make sure that the light curves have a good signal-to-noise ratio for the analysis. \

Finally, we crossmatched our list with the \textit{Gaia} DR3 catalogue \citep{GaiaDR3summary}, which served multiple purposes. One was to provide a universal ID for all our stars that we can use consistently. Another was to check the \textit{Gaia} RR Lyrae catalogue \citep{Clementini}, too, which contained valuable information about 89\% of our RR Lyraes, such as the type and pulsation period in the fundamental mode and/or the first overtone. Having another point of comparison in terms of classification, the pulsation period is important, especially in cases in which there is a disagreement between the Pan-STARRS and K2 observations in any way. \

\begin{figure*}
    \includegraphics[width=\textwidth]{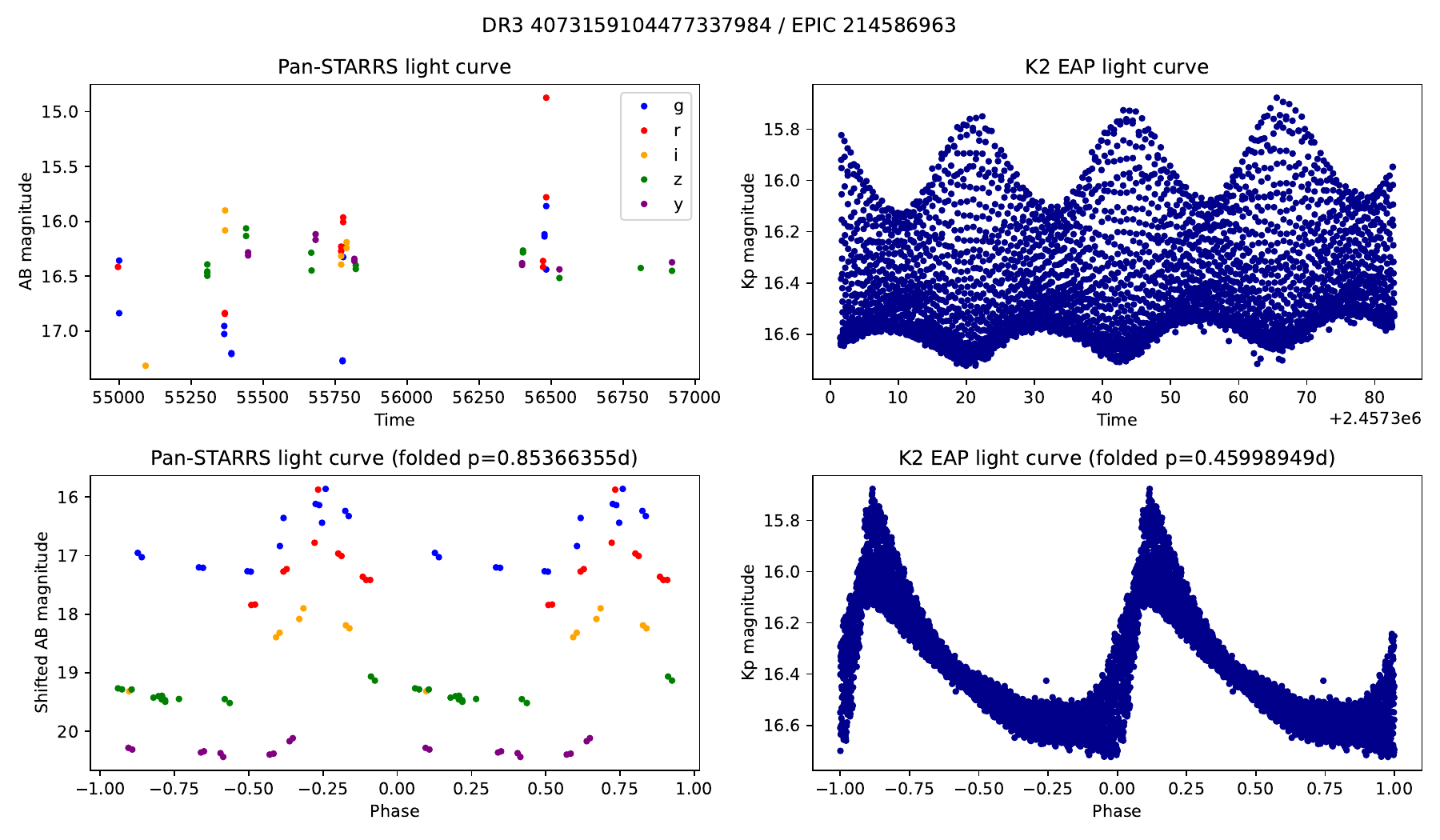}
    \caption{Light curve comparison for a sample RR Lyrae star. The upper left panel visualises the Pan-STARRS light curve of the star, in all five filters, and the lower left panel shows the light curve phase-folded with the period given in the \citep{Sesar2017} catalogue. In order to make the periodicity more transparent, we shifted the AB magnitude levels of each filter's light curve by adding 0,1,2,3,4 magnitudes to the g,r,i,z,y filter light curves. The upper right panel illustrates the K2 light curve, while the lower right panel shows the light curve phase-folded with the period that was determined by the Lomb-Scargle periodogram method.}
    \label{fig:lc_compare}
\end{figure*}

For all of the stars in the \citet{Sesar2017} catalogue, we refer to their provided period as the Pan-STARRS period. We calculated the K2 periods from the autoEAP light curves using the Lomb-Scargle periodogram method in the Astropy Python package \citep{astropy:2022}. However, in Campaign 7 of K2, many light curves are contaminated due to the dense field and we found that Pre-search Data Conditioning Simple Aperture Photometry (PDCSAP, \citep{VanCleve2016}) produced a cleaner light curve. Therefore, for these affected stars, PDCSAP photometry was used instead of the autoEAP one. For the stars that are found in the \textit{Gaia} DR3 RR Lyrae catalogue \citep{Clementini}, the \textit{Gaia} frequency is also given. \

For further analysis, we used the corrected autoEAP light curves from K2 and downloaded the detections in each filter from Pan-STARRS. This step served only a qualitative purpose; the two light curves were visually compared. To be able to make a comparison, we converted the light curves to magnitude units. The magnitude conversion formula of $\mathrm K_p=-2.5\,log_{10}(\mathrm{F})+25.3$, calibrated by \citet{lund2015}, was applied to convert the K2 flux to a $\mathrm K_p$ magnitude. We used the $\mathrm m_{AB}=-2.5\,log_{10}(\mathrm{F}/3631 Jy)$ formula \citep{Tonry2012} to convert the Pan-STARRS light curves from Jansky to the AB magnitude system for each passband. \

Finally, we checked the K2 light curves of all the variable stars on the final list and classified them by visual inspection. Light curves were classified based on the periods and the structure of the folded light curves, and were compared to templates, such as the OGLE Atlas of Variable Stars,\footnote{\url{https://ogle.astrouw.edu.pl/atlas/index.html}} if needed. The \citet{Sesar2017} catalogue lists their classifications based on the Pan-STARRS data; however, the only two subtypes in the catalogue are RRab (fundamental mode pulsator) and RRc (first overtone pulsator) stars. RRd (double mode pulsator) stars or other variable types were not present as classification categories. Therefore, it was crucial to obtain the K2-based classification of the stars as well, to be able to better characterise and analyse the sample. \

\section{Results} \label{results}

The main properties of the stars that our search criteria yielded are listed in Table\textit{~\ref{table:1}} in the appendix. The entire sample consists of 1120 RRab and 233 RRc stars, classified by \citet{Sesar2017}. The whole table is available online.\footnote{\url{https://kik.konkoly.hu/data_en.html}} The first two columns list the co-ordinates of the stars (taken from the \citet{Sesar2017} catalogue), the second two columns the EPIC (K2) and \textit{Gaia} DR3 IDs, the fifth column is the classification type by \citet{Sesar2017}, and the sixth column is the classification type based on the K2 light curve. Then, the seventh, eighth, and ninth columns show the pulsation frequencies of the RR Lyraes. The K2 ones were calculated using the Lomb-Scargle periodogram method; meanwhile, the Pan-STARRS frequencies were taken from the \citet{Sesar2017} catalogue and the \textit{Gaia} frequencies are from the DR3 Special Object Studies (SOS) RR Lyrae variable catalogue \citep{Clementini}. The last two columns show the $\mathrm K_p $ taken from the EPIC catalogue \citep{Huber2016} and the flux-averaged g (<g>) magnitudes taken from the \citet{Sesar2017} catalogue.\

To illustrate a typical case, the light curves of one of the sample RR Lyrae stars are shown in \textit{Figure~\ref{fig:lc_compare}}. It is clear that in all cases the Pan-STARRS light curves are sparsely sampled compared to the K2 light curves; however, their longer observation interval leads to more precise periods. One important thing to note in this example is that the period that was calculated based on the K2 light curve using the Lomb-Scargle periodogram method is different from that determined by \citet{Sesar2017}. \

In the case of the K2 data, the identification (and period determination) of the RR Lyrae star is straightforward and can be done with good precision based on the time series and the folded light curves, even by visual inspection. However, as can be seen in the left-hand panels, the Pan-STARRS datasets contain significantly fewer detections, making it harder to track the brightness changes through the different phases. The catalogue of RR Lyrae stars in the Pan-STARRS data was compiled based on template light curve fits using machine-learning \citep{Sesar2017}. The corresponding template was not included in \textit{Figure~\ref{fig:lc_compare}} because one of the free parameters (r') could not be recovered from the work of \citet{Sesar2010}.  \

\begin{figure*}
    \includegraphics[width=\textwidth]{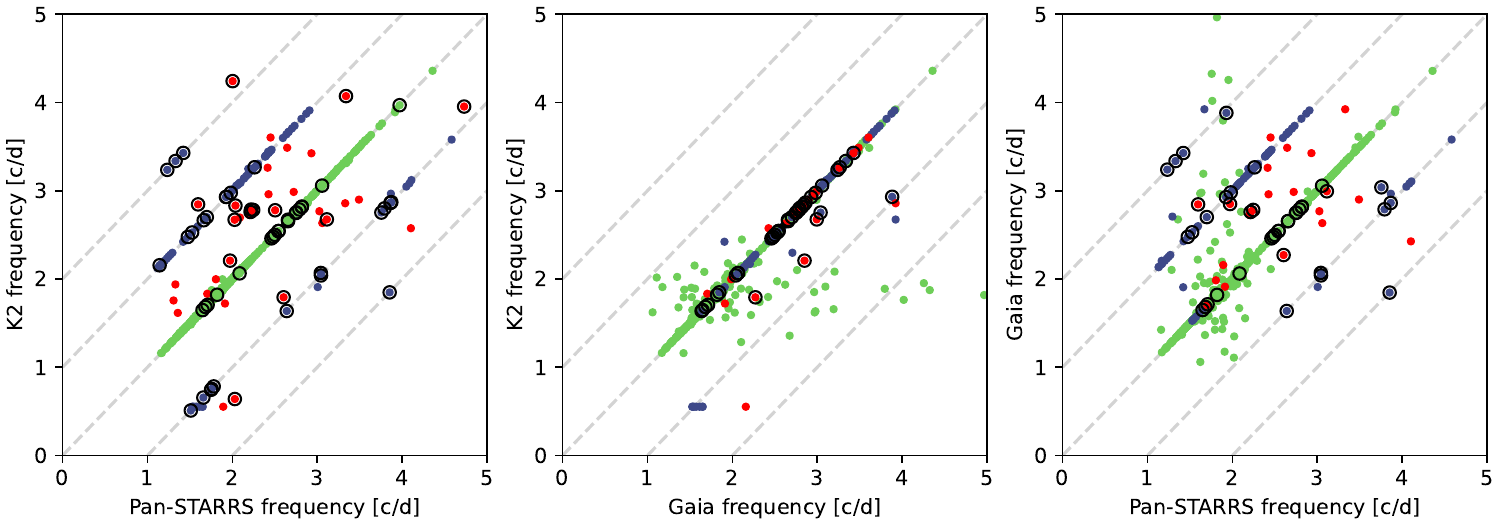}
    \caption{Frequency comparisons of the sample. The figure shows the comparisons of the periods determined from different datasets. For Pan-STARRS, the \citet{Sesar2017} catalogue period was used, for \textit{Gaia} DR3, the \textit{Gaia} RR Lyrae catalogue period was used, and for K2, our calculated values were used, which we obtained with the Lomb-Scargle periodogram method. The plot symbols are colour-coded based on the Pan-STARRS vs K2 frequency comparison. Green dots indicate the stars that have matching frequencies, blue dots indicate the stars that have an alias of 1 or 2 cycles/day in their Pan-STARRS frequencies, and red dots signify the stars that have discrepancies between their K2 and Pan-STARRS frequencies that are not aliases. Black circles mark the stars that we found to be misclassified in the \citet{Sesar2017} catalogue.}
    \label{fig:periods}
\end{figure*}

\begin{figure*}
    \includegraphics[width=\textwidth]{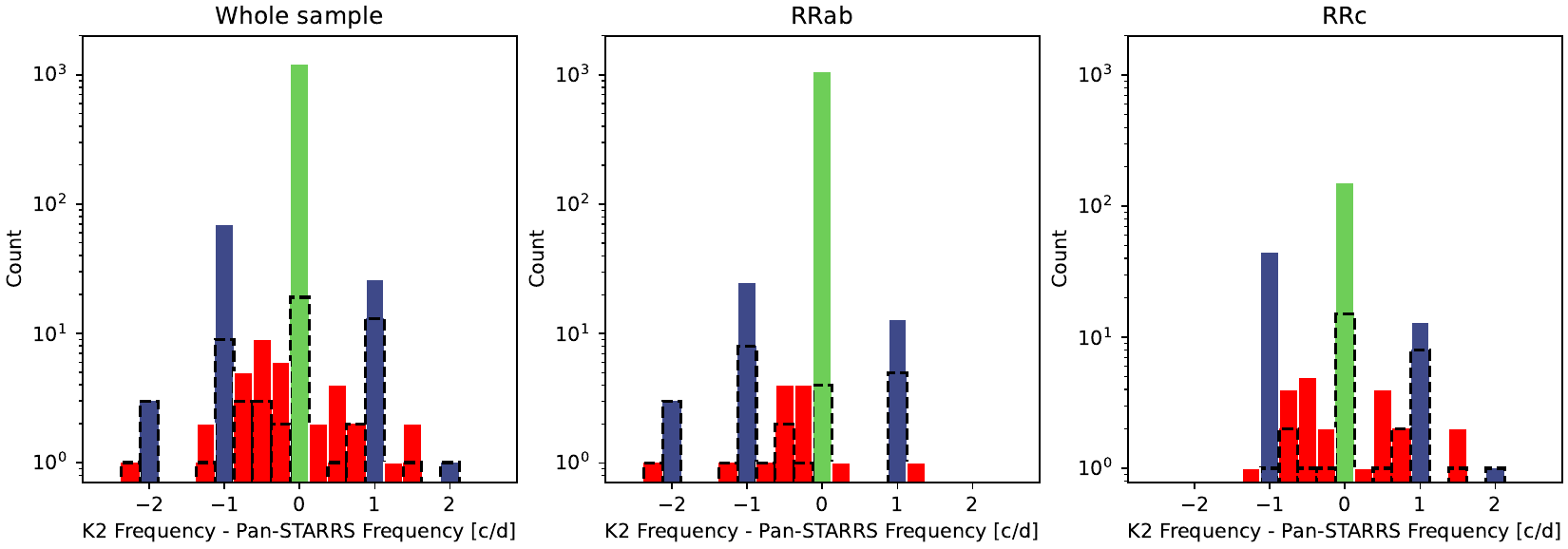}
    \caption{Distibution of frequency differences for each sub-type of RR Lyrae stars. The classification types were taken from the \citet{Sesar2017} catalogue. Green indicates the stars that have matching frequencies, blue indicates the stars that have an alias of 1 or 2 cycles/day in their Pan-STARRS frequencies, and red signifies the stars that have discrepancies between their K2 and Pan-STARRS frequencies that are not aliases. The dot-edged black columns show the number of misclassified stars within each bin.}
    \label{fig:hist}
\end{figure*}

\begin{figure*}
    \includegraphics[width=\textwidth]{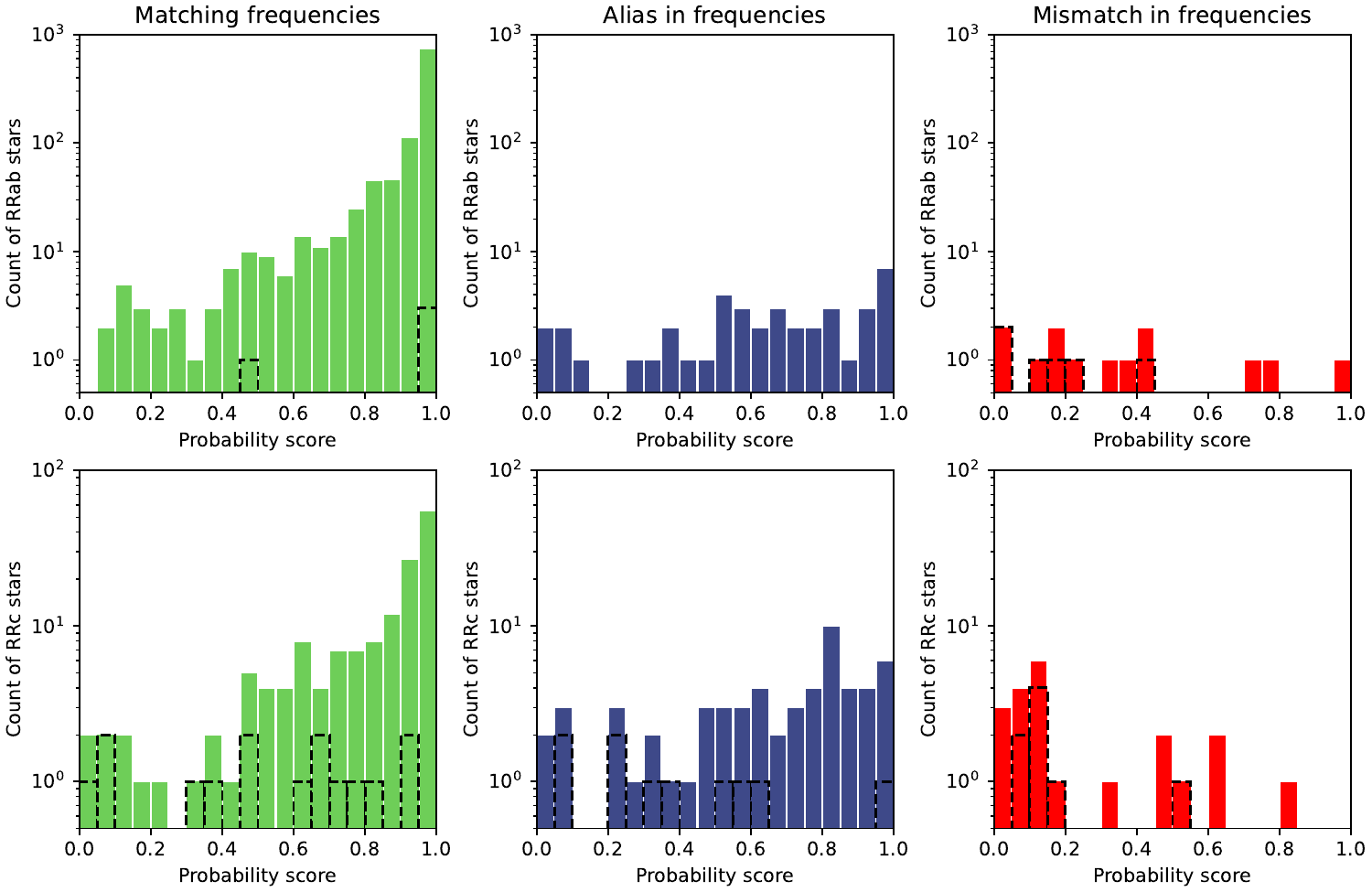}
    \caption{RR Lyrae classification probability score distribution of the crossmatched sample. The classification probability score of each star was given in the \citet{Sesar2017} catalogue. The color green refers to the stars with matching frequencies from K2 and Pan-STARRS, blue denotes stars that have an alias in their frequency from Pan-STARRS, and red is for the stars that have a difference in their frequencies. The dot-edged black columns show the number of misclassified stars within each bin.}
    \label{fig:scores}
\end{figure*}

For the majority of the stars (90.1\% in total, 95.2\% of RRab stars, and 65.7\% of RRc ones), the periods calculated from the K2 light curves are consistent with those listed in the \citet{Sesar2017} catalogue within the uncertainties. We estimated the uncertainties from the frequency resolution of the data. The resolution of a frequency spectrum can be approximated  with $\Delta f=2/T$ \citep{Aerts_book}. Assuming that the average length of a K2 light curve is $T=80$ days, the formula yields $ \Delta f=0.025\,d^{-1}$. We estimated the uncertainties of the Pan-STARRS observations similarly; that is, of the order of $ \Delta f=0.001\,d^{-1}$. Frequency uncertainties were published in the \textit{Gaia} DR3 catalogue. A comparison with the \textit{Gaia} data indicates that our estimates serve as upper limits for the frequency estimates, but more accurate values would require calculations on a star-by-star basis. \ 

For a significant subset of the sample (9.9\% in total, 4.8\% of RRab stars, and 34.3\% of RRc ones), the period calculations yielded quite different results, as is demonstrated in the case of EPIC 214586963 (\textit{Figure~\ref{fig:lc_compare}}), in which the frequencies for the star differ by 1~$d^{-1}$. We managed to determine a clear pattern in the frequency differences, which is shown in \textit{Figure~\ref{fig:periods}}. For 7.4\% of the whole sample (3.7\% of RRab and 25.3\% of RRc stars) we found an alias in the frequencies, meaning that the frequencies given in the \citet{Sesar2017} catalogue are 1.0 or 2.0~$d^{-1}$ off compared to the frequencies derived from the K2 light curves. It is clearly visible that there is an asymmetry in the alias frequencies; the frequencies determined from the Pan-STARRS light curves are more frequently 1.0~$d^{-1}$ shorter than longer. For the remaining 2.5\% of stars (1.2\% of RRab and 9.0\% of RRc stars), we found discrepancies between the two frequencies, yet no aliases. For the outlier stars that also have \textit{Gaia} DR3 variability data, in 63.6\% of cases the frequency given by \textit{Gaia} is in agreement with the K2 frequencies. \

\begin{table*}
\caption{Summary of the result of the frequency analysis.}\label{table:2}
\centering
\begin{tabular}{c c c c c c c c}
\hline\hline             
Variables & Final sample & Match & Match & Alias & Alias & Mismatch & Mismatch \\ 
\hline\hline
RRab & 1120 [1094] & 1066 [1062] & 95.2 \% & 41 [25] & 3.7 \% & 13 [7] & 1.2 \% \\
RRc & 233 [200] & 153 [138] & 65.7 \% & 59 [49] & 25.3 \% & 21 [13] & 9.0 \% \\
Total & 1353 [1294] & 1219 [1200] & 90.1 \% & 100 [74] & 7.4 \% & 34 [20] & 2.5 \% \\
\hline
\end{tabular}
\tablefoot{The second column lists how many stars the sample had of each type of RR Lyrae star. The third and fourth columns show the number and percentage of stars that had matching frequencies from both Pan-STARRS and K2. The fifth and sixth columns show the number and percentage of stars that had an alias of 1 or 2 cycles/day in their Pan-STARRS frequencies compared to the ones obtained from K2. Finally, the seventh and eighth columns show the number and percentage of stars that had different frequencies (not aliases) from K2 and Pan-STARRS. The numbers in brackets indicate the number of stars within each group for which the classification is consistent.}
\end{table*}

\begin{figure}
    \includegraphics[width=\columnwidth]{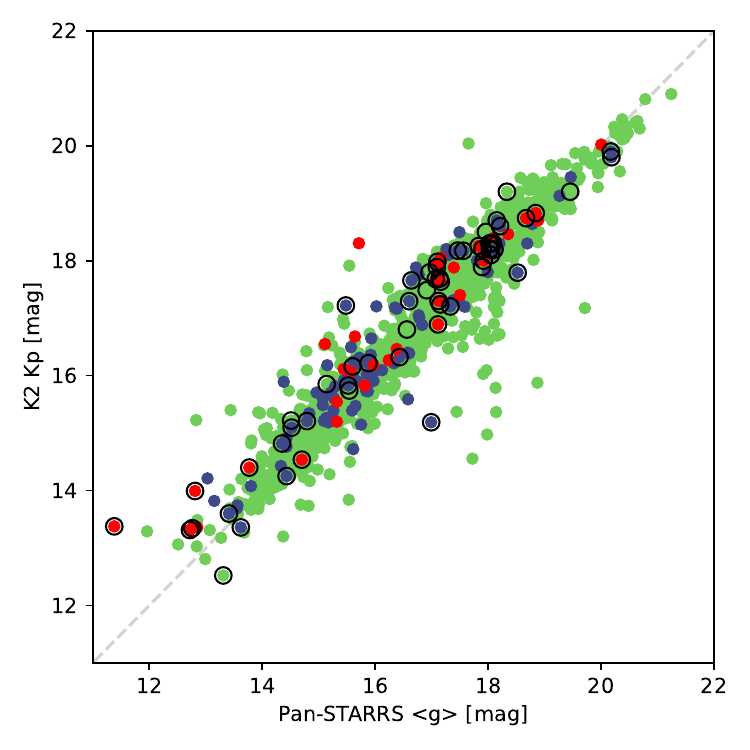}
    \caption{Brightness comparison of the sample. Pan-STARRS \textit{g} magnitudes are shown on the horizontal axis, while \textit{Kepler} \textit{Kp} magnitudes are plotted for each star on the vertical axis. The green dots denote stars that have the same frequency from both K2 and Pan-STARRS data, the blue dots are stars that have an alias of 1 or 2 $d^{-1}$ in their Pan-STARRS frequencies, and red dots mark the stars for which there is a different difference in the periods. Black circles mark the stars that we found to be misclassified in the \citep{Sesar2017} catalogue.}
    \label{fig:mag}
\end{figure}

In \textit{Figure~\ref{fig:periods}}, we compare the identified frequencies for each star using K2, Pan-STARRS, and even \textit{Gaia} DR3 data when available. We found that when there is a mismatch in the frequencies, it is likely to be off by 1.0 or 2.0 $d^{-1}$, the 1.0 $d^{-1}$ offset being the most common one. This trend is observable in both Pan-STARRS versus K2 and Pan-STARRS versus \textit{Gaia} comparisons, but not in the K2 versus \textit{Gaia} analysis. Therefore, for stars with discrepant period values, we recommend using the K2 or \textit{Gaia} periods instead of the Pan-STARRS values.

The explanation for this finding is the biases present in the different kinds of observations. K2 and \textit{Gaia} are space telescopes, able to make observations both during daytime and nighttime, but Pan-STARRS, being a ground-based telescope, could observe only during local nighttime. These regularly missing intervals from the sampling are likely the causes of this phenomenon. \

We analysed not only the frequencies of the variables, but also their classifications. We compared the classification given in the \citet{Sesar2017} catalogue (RRab or RRc) with the classification that resulted from the visual inspection of the K2 light curves. We found that for 95.6 \% of the stars (97.7 \% of RRab stars and 85.8 \% of RRc stars), the classifications were consistent. In five cases, classification based on the K2 light curve was not possible due to the quality of the light curve. With these exceptions, we are confident that the K2-based classification is the correct one. Among the stars that were misclassified, we found examples where, although the star was an RR Lyrae type star, the correct subtype was different; for example, RRd instead of RRab or RRc instead of RRab and vice versa. We found \textbf{12} cases in which based on the K2 light curves the star is clearly not an RR Lyrae star, but rather another type of variable, such as BL~Herculis, $\delta$~Scuti, an anomalous Cepheid star, or an eclipsing binary. In total, 29.6 \% of the misclassified stars have matching frequencies, 46.3 \% have aliases, and 24.1~\% have different frequencies from K2 and Pan-STARRRS. The results are summarized in \textit{Table~\ref{table:2}}. \

The distribution of the frequency differences is shown in \textit{Figure~\ref{fig:hist}}. From the second and third panels of the plot, it is obvious that RRc stars are disproportionately affected. This could be attributed to the fact that the RRc light curve shape is more sinusoidal. Furthermore, the fits done by \citet{Sesar2017} find frequencies lower by 1.0 $d^{-1}$ significantly more often. This suggests that assumptions made for the preferred frequency range could also have affected the results, preferring lower frequency fits for those stars. When considering the classifications, it is important to note that we find misclassified stars in all three groups, not just in the period mismatch group, which indicates that just because the classification is not accurate, the given frequency can be. \

The distribution of the probability scores of the overlapping sample stars is presented in \textit{Figure \ref{fig:scores}}. These scores are given in the \citet{Sesar2017} catalogue and are indicators of their RR Lyrae classification quality. As is apparent from the semi-log scale charts, the vast majority of the stars have a probability score at the higher end for each sub-class. Although in the case of the stars with matching periods there does not seem to be a relation between the probability score values and the number of misclassified stars, in the case of the alias group and especially in the period mismatch group, as was expected, we tend to find more misclassified stars among those that have lower-probability score values in the \cite{Sesar2017} catalogue. This effect is particularly noticeable in case of the RRc stars. \

Finally, we compare the brightnesses of the variables in \textit{Figure~\ref{fig:mag}}. From the Pan-STARRS datasets, we used the \textit{g}-filter data to compare with the K2 {$\mathrm K_p$} magnitudes. The brightnesses derived from the K2 and the Pan-STARRS light curves are in agreement for the majority of the stars in the sample; however, there are cases with significant differences in the magnitudes. There is no visible link between the correctness of the period or classification of the stars and the brightness differences. The discrepancies are rather due to the uncertainties in the determination of the brightnesses. \

\section{Summary and conclusions}
\label{conclusions}

The Pan-STARRS 3$\pi$ survey has detected a vast number of RR~Lyrae stars, but these light curves are sparsely sampled, and the observations were only possible during local nighttime as a result of it being a ground-based telescope. On the other hand, the K2 mission only observed a limited fraction of the sky in each of its campaigns, but did so with a much higher sampling frequency, and thanks to the fact that it was a space mission, the time series are quasi-continuous. Their data can complement each other: \textit{Kepler} provides high-time-resolution light curves for a shorter period of time for a limited number of stars, whereas Pan-STARRS provides more sparsely sampled data but for a longer time span and for significantly more stars. Because of this, we decided to investigate their overlap. To carry out a more thorough analysis, we used the \textit{Gaia} DR3 data as well, wherever it was available. \

For the comparison, we derived the classification and period information from the \citet{Sesar2017} catalogue. From the K2 data, we used the available autoEAP light curves and calculated the periods with the Lomb-Scargle periodogram method. We found that for the majority of the stars (95.6 \%), the classification was consistent, but we also found some exceptions; for example, some RRd stars were classified as RRab or RRc stars, or the stars were other type of pulsators or eclipsing binaries instead. \

During the investigation of the periods, we found that for the majority (90.1\%) of the stars, the periods derived from the Pan-STARRS and K2 data are consistent. However, we found a significant subset of stars (9.9\%) that seem to have different periods from the two missions. \

Analysing the frequencies, it became apparent that there is a systematic difference between the periods calculated from K2 and those from the catalogue in the case of some stars (7.4\% of the sample). We found that the Pan-STARRS frequencies were usually 1 or 2 $d^{-1}$ off from the K2 periods. This discrepancy strongly affected the RRc periods, with a quarter of them (25.3\%) being alias values instead of the true ones. The explanation of this alias is the fact that because Pan-STARRS observed from the ground, and therefore could only make measurements during nighttime, the sampling is not even, compared to the time series of the K2 mission. The \textit{Gaia} periods (where available) are in agreement with the K2 periods in most cases (94.6\%). In terms of the remainder, for 3.33\% of the stars only the \textit{Gaia} frequency is off and the K2 and Pan-STARRS frequencies are in agreement, and for 1.48\% of the stars all three frequencies are different. \

Our research concludes that although Pan-STARRS observed a huge number of stars with measurements spanning a long time interval (four years), it was still important and insightful to compare the data with the K2 mission data, not only because K2 offers a more evenly and frequently sampled light curve, but more importantly because there are aliases in the determined frequencies due to the observing time bias of ground-based detections and these aliases are not necessarily symmetrical. In our case it was more likely to find the Pan-STARRS frequency \citep{Sesar2017} to be 1 $d^{-1}$ lower than higher than the K2 frequency. \

The conclusions of the study are important in the era of big surveys such as the \textit{Vera C. Rubin Observatory}'s LSST (Legacy Survey of Space and Time) sky survey \citep{LSST}. The LSST program  will produce an unprecedented amount of data on the stars in the southern hemisphere for ten years with a relatively sparse cadence. Therefore, validating the ground-based data against space-based observations is critical in being able to identify the possible aliases and other artefacts in the dataset that can later be taken into account during data reduction in order to get more precise and accurate results.\

\begin{acknowledgements}
This project has been supported by the Lend\"{u}let Program of the Hungarian Academy of Sciences, project No. LP2018-7/2022, the KKP-137523 'SeismoLab' \'{E}lvonal and the NKFIH SNN-147362 grants of the Hungarian Research, Development and Innovation Office (NKFIH).  This work was also supported by the NKFIH excellence grant TKP2021-NKTA-64.
This paper includes data collected by the \textit{Kepler} mission and obtained from the MAST data archive at the Space Telescope Science Institute (STScI). Funding for the \textit{Kepler} mission is provided by the NASA Science Mission Directorate. STScI is operated by the Association of Universities for Research in Astronomy, Inc., under NASA contract NAS 5–26555.
The Pan-STARRS1 Surveys (PS1) and the PS1 public science archive have been made possible through contributions by the Institute for Astronomy, the University of Hawaii, the Pan-STARRS Project Office, the Max-Planck Society and its participating institutes, the Max Planck Institute for Astronomy, Heidelberg and the Max Planck Institute for Extraterrestrial Physics, Garching, The Johns Hopkins University, Durham University, the University of Edinburgh, the Queen's University Belfast, the Harvard-Smithsonian Center for Astrophysics, the Las Cumbres Observatory Global Telescope Network Incorporated, the National Central University of Taiwan, the Space Telescope Science Institute, the National Aeronautics and Space Administration under Grant No.\ NNX08AR22G issued through the Planetary Science Division of the NASA Science Mission Directorate, the National Science Foundation Grant No.\ AST-1238877, the University of Maryland, Eötvös Loránd University (ELTE), the Los Alamos National Laboratory, and the Gordon and Betty Moore Foundation.
This work has made use of data from the European Space Agency (ESA) mission
{\it Gaia} (\url{https://www.cosmos.esa.int/gaia}), processed by the {\it Gaia}
Data Processing and Analysis Consortium (DPAC,
\url{https://www.cosmos.esa.int/web/gaia/dpac/consortium}). Funding for the DPAC
has been provided by national institutions, in particular the institutions
participating in the {\it Gaia} Multilateral Agreement.
\end{acknowledgements}

\bibliographystyle{aa}
\bibliography{example}

\begin{appendix} 
\section{Table of investigated RR Lyrae variables}
\begin{sidewaystable*}
\caption{The main properties of the investigated RR Lyrae variables.}\label{table:1}
\centering
\begin{tabular}{c c c c c c c c c c c c}
\hline\hline             
RA & Dec & EPIC ID & Gaia DR3 ID & Type & Type & $\mathrm F_{Pan-STARRS}$ & $\mathrm F_{K2}$ & $\mathrm F_{Gaia}$ &  $\mathrm K_p$ & <g> & Frequency status \\ 
deg & deg & - & - & Pan-STARRS & K2  & c/d & c/d & c/d & mag & mag & - \\
\hline\hline
9.63422 & 3.50196 & 220340933 & 2553727016638405760 & RRab & RRab & 1.924141 & 1.9229 & 1.924103 & 15.61 & 15.60 & Match\\
10.33102 & 5.34633 & 220432468 & 2554575083700694656 & RRab & RRab & 1.918994 & 1.9179 & 1.919098 & 13.81 & 13.85 & Match\\
10.83601 & 2.37782 & 220287868 & 2550359830997925120 & RRab & RRc & 1.698565 & 2.7005 & 2.701171 & 17.20 & 17.33 & Alias\\
11.59281 & 2.92901 & 220313822 & 2550769742676812032 & RRab & RRab & 2.017497 & 2.0170 & 2.017539 & 17.17 & 16.84 & Match\\
12.26802 & 2.42101 & 220289837 & 2549943734566154368 & RRab & RRab & 1.899221 & 1.9001 & 1.899147 & 13.92 & 14.01 & Match\\
12.34864 & -0.29986 & 220192221 & 2536781236056128640 & RRab & RRab & 1.938215 & 1.9382 & 1.938196 & 18.41 & 17.54 & Match\\
12.39597 & 2.38178 & 220288040 & 2549930712225323520 & RRab & RRab & 1.842823 & 1.8416 & 1.842870 & 14.08 & 14.28 & Match\\
12.59608 & 3.88395 & 220359154 & 2552435739015223296 & RRab & RRab & 1.989808 & 1.9890 & 1.989829 & 16.67 & 17.18 & Match\\
13.22465 & -0.36159 & 229228813 & 2536811781863572736 & RRab & RRab & 1.723772 & 1.7247 & 1.723777 & 18.70 & 19.13 & Match\\
13.36934 & -0.72254 & 220181683 & 2535946638011222144 & RRab & RRab & 1.816330 & 1.8162 & 1.816426 & 18.02 & 17.58 & Match\\
13.40920 & -0.05072 & 220198696 & 2536827282401261568 & RRab & RRab & 1.897475 & 1.8975 & 1.897423 & 16.06 & 16.53 & Match\\
13.53985 & 0.48317 & 229228816 & 2537303607158516608 & RRab & RRab & 1.554635 & 1.5545 & 1.554218 & 20.30 & 20.68 & Match\\
13.55268 & 2.81758 & 220308445 & 2549864118756805248 & RRab & RRab & 1.545228 & 1.5443 & 1.545356 & 19.08 & 18.94 & Match\\
13.68262 & 4.37956 & 220383290 & 2551815137716234880 & RRab & RRab & 1.985449 & 1.9865 & 1.985495 & 16.67 & 16.63 & Match\\
13.85890 & 0.88852 & 229228817 & 2537343674908906496 & RRab & RRab & 1.933713 & 1.9331 & N/A & 20.40 & 20.60 & Match\\
14.30116 & -0.09931 & 229228806 & 2536447809155139328 & RRc & RRc & 1.595431 & 2.5989 & 2.598104 & 17.80 & 17.99 & Alias\\
14.49476 & 7.04089 & 220511999 & 2577540578774756992 & RRab & RRab & 1.558973 & 1.5595 & 1.558771 & 18.38 & 18.10 & Match\\
14.50523 & 12.09602 & 220721216 & 2584307488728353024 & RRab & RRab & 1.915702 & 1.9153 & 1.915702 & 17.21 & 16.93 & Match\\
14.61663 & -0.75059 & 220180957 & 2535805045825145344 & RRc & RRc & 1.997468 & 3.0004 & 3.000255 & 18.20 & 17.26 & Alias\\
14.75640 & 1.41287 & 220242787 & 2537469599055061248 & RRab & RRab & 1.791780 & 1.7908 & 1.791838 & 18.98 & 18.86 & Match\\
14.79055 & 3.55358 & 220343343 & 2551464805823863680 & RRab & RRab & 1.775260 & 1.7755 & 1.775270 & 16.11 & 15.68 & Match\\
14.79961 & 7.40232 & 220528452 & 2577579375214836736 & RRab & RRab & 1.795416 & 1.7959 & 2.420274 & 17.65 & 16.83 & Match\\
15.08583 & 5.19342 & 220424589 & 2552346712933033088 & RRab & RRab & 1.681237 & 1.6815 & 1.681258 & 17.35 & 17.22 & Match\\
15.33073 & 1.63805 & 220253247 & 2538936489300611840 & RRab & RRab & 2.083017 & 2.0830 & 2.083023 & 17.71 & 17.51 & Match\\
15.62110 & -0.05924 & 220198467 & 2536319582906997248 & RRab & RRab & 1.573670 & 1.5748 & 1.573673 & 15.82 & 15.83 & Match\\
15.99581 & 4.28111 & 220378583 & 2551904473036022784 & RRab & RRab & 1.738133 & 1.7374 & N/A & 15.80 & 15.80 & Match\\
16.05539 & 0.45371 & 229228819 & N/A & RRab & RRab & 1.952408 & 1.9534 & N/A & 20.90 & 21.24 & Match\\
16.11449 & 11.14098 & 220688027 & 2583222992305882112 & RRc & RRc & 2.795808 & 2.7971 & 2.795829 & 17.83 & 17.88 & Match\\
16.20788 & 5.28936 & 220429650 & 2552132037583250560 & RRab & RRab & 1.454882 & 1.4554 & 1.454925 & 15.15 & 15.31 & Match\\
16.37094 & -1.50281 & 220161256 & 2533005689220708224 & RRab & RRab & 1.436525 & 1.4376 & 1.436513 & 14.07 & 14.03 & Match\\
16.39052 & -0.39567 & 229228810 & 2536239623500251776 & RRab & RRab & 1.594121 & 1.5951 & 1.594132 & 17.70 & 17.93 & Match\\
16.49875 & -0.37890 & 220190195 & 2533238648246867200 & RRab & RRab & 1.701496 & 1.7018 & 1.701591 & 16.74 & 16.83 & Match\\
16.50138 & -0.30943 & 220191991 & 2536241964258088320 & RRab & RRab & 1.697782 & 1.6968 & 1.697792 & 14.19 & 14.14 & Match\\
16.68210 & 8.49499 & 220575441 & 2578081847732991104 & RRab & RRab & 2.143079 & 2.1440 & 2.143100 & 17.58 & 17.78 & Match\\
16.76019 & 6.53570 & 220489863 & 2576562838059567104 & RRab & RRab & 1.672978 & 1.6739 & N/A & 17.70 & 17.06 & Match\\
17.29446 & 7.19756 & 220518877 & 2576988314699581056 & RRab & RRab & 1.877968 & 1.8772 & N/A & 19.45 & 19.14 & Match\\
17.32055 & -2.16368 & 220142950 & 2531992008219298432 & RRc & RRc & 3.574803 & 3.5747 & N/A & 14.98 & 14.78 & Match\\
17.67746 & 9.28213 & 220611561 & 2579732184622054144 & RRc & RRc & 2.782734 & 2.7819 & 2.782648 & 16.56 & 16.52 & Match\\
17.69362 & -2.03746 & 220146083 & 2531989530022524032 & RRab & RRab & 1.713834 & 1.7146 & 1.713658 & 17.61 & 17.57 & Match\\
\hline
\end{tabular}
\tablefoot{The first two columns of the table contain the co-ordinates of the stars taken from the \citet{Sesar2017}, the third lists the K2 EPIC ID, the fourth is the Gaia DR3 ID. The fifth column shows the classification given by the \citet{Sesar2017} catalogue, the sixth column contains the classification we provide using the K2 light curves. The seventh column lists the frequencies from the \citet{Sesar2017} catalogue, eighth column shows the frequencies that were calculated from the K2 EAP light curves. The ninth column shows the Gaia frequencies. The tenth column contains the $\mathrm K_p$ magnitude values that were taken from the EPIC catalogue \citep{Huber2016} and the eleventh column contains the flux-averaged g-magnitude values as derived from the \citet{Sesar2017} catalogue. The final column shows whether the frequencies from K2 and Pan-STARRS were in agreement. This table only contains the first 40 records of the whole sample, the entire table is available online (\url{https://kik.konkoly.hu/data_en.html}).}
\end{sidewaystable*}
\end{appendix}

\end{document}